\documentclass{desyproc}

\usepackage{subfigure}
\usepackage{axodraw4j}
\usepackage{cite}

\newcommand{\Aslash}{A\!\!\!/\,}
\newcommand{\Bslash}{X\!\!\!\!/\,}

\begin{document}
\title{\vspace{-3cm}
\hfill{\small{IPPP/09/60; DCPT/09/120}}\\[2cm]
Strong fields and recycled accelerator parts as a laboratory for
fundamental physics\\
--\\
\large{The quest for minicharged particles}
}

\author{{\slshape Joerg Jaeckel}\\[1ex]
Institute for Particle Physics and Phenomenology, Durham University, Durham DH1 3LE, UK\\
}

\contribID{jaeckel\_joerg}

\maketitle

\begin{abstract}
Over the last few years it has become increasingly clear that low energy, but high precision experiments provide a powerful and complementary
window to physics beyond the Standard Model.
In this note we illuminate this by using minicharged particles as an example. We argue that minicharged particles arise naturally
in extensions of the Standard Model. Compatibility with charge quantization arguments suggests that minicharged particles typically arise together
with a massless hidden sector U(1) gauge field. We present several low energy experiments employing strong lasers, electric and magnetic fields that can be used
to search for (light) minicharged particles and their accompanying U(1) gauge boson.

\end{abstract}

\section{Introduction}
The development of high energy accelerators also leads to the development of technology to generate high electric and magnetic fields.
In this note we want to explore how one can directly use these strong fields as laboratory for fundamental physics.
High precision experiments in these strong fields allow us to search for new light particles which interact very
weakly with the Standard Model particles.
Due to their extremely weak interaction these hidden sector particles may be missed in conventional collider experiments.
Accordingly the discussed high precision experiments provide complementary information on physics beyond the Standard Model.

In this note we will mainly discuss a particular type of hidden sector particle: minicharged particles.
However, as we will see minicharged particles are often accompanied by an extra U(1) gauge boson, a hidden photon.

Minicharged particles interact with the ordinary electromagnetic field via the usual minimal coupling induced by the covariant derivative,
\begin{equation}
D_{\mu}=\partial_{\mu}-{\mathbf{i}}Q_{f}eA_{\mu}
\end{equation}
where $Q_{f}$ is the electric charge of the particle $f$.
For example if $f$ is a fermion the interaction term reads
$Q_{f}e \bar{f}A\!\!\! / f.$

The crucial point for a minicharged particle is now simply that the charge is much smaller than~$1$,
\begin{equation}
\epsilon=Q_{f}\ll 1.
\end{equation}
In particular it is not necessarily integer. Indeed it does not even have to be a rational number.
Minicharges can arise in theories with kinetic mixing \cite{Holdom:1985ag}, or magnetic mixing~\cite{Bruemmer:2009ky}
but also in scenarios with extra dimensions \cite{Batell:2005wa}.
Typical predicted values, e.g., in realistic string compactifications range from $10^{-16}$ to $10^{-2}$ \cite{Batell:2005wa,Dienes:1996zr}.

\section{Minicharged particles in extensions of the Standard Model}
Before we investigate how minicharges can be experimentally searched for in strong field experiments let us briefly show how they
arise in simple extensions of the Standard Model. As we will see, in this simple setup the minicharged particles are accompanied
by an extra U(1) gauge boson which also opens up new possibilities for experimental searches.

One way to have minicharged particles is in extensions of the Standard Model that contain
an extra hidden U(1) gauge degree of freedom under which the Standard Model particles are uncharged.
Such a model has a gauge group U$(1)_{\rm QED}\times$ U$(1)_{h}$.
The most general renormalizable Lagrangian in the gauge sector is,
\begin{equation}
\label{lagrangian}
\begin{split}
{\cal L}=-\frac{1}{4}\Bigl\{\,&\frac{1}{e^2}\,F^{\mu\nu} F_{\mu\nu}\,
+\,\frac{1}{e_h^2}\,X^{\mu\nu} X_{\mu\nu}\,+\,\frac{2\chi}{e e_h}\, F^{\mu\nu}X_{\mu\nu}\\
&+\,\frac{1}{8\pi^2}\left(\theta_{\rm QED}\,F^{\mu\nu}\tilde F_{\mu\nu}\,+\,\theta_{h}\,X^{\mu\nu}\tilde X_{\mu\nu}\,
+\,2\theta_{\chi}\,F^{\mu\nu}\tilde X_{\mu\nu}\right)\Bigr\}.
\end{split}
\end{equation}
Here, the $F^{\mu\nu}$ is the field strength of the electromagnetic U(1)$_{\rm QED}$ with the gauge field $A^{\mu}$ and
$X^{\mu\nu}$ is the field strength of the hidden U(1)$_{h}$ with the gauge field $X^{\mu}$.
The gauge couplings are $e$ and $e_{h}$, respectively.
The dual field strengths are defined as usual:
\begin{equation}
\tilde F^{\mu\nu}=\frac{1}{2}\epsilon^{\mu\nu\kappa\lambda}F_{\kappa\lambda}\,,
\end{equation}
and analogously for $\tilde{X}^{\mu\nu}$.

The standard kinetic terms as well as the
$\theta_{\rm QED}$- and $\theta_{h}$-terms are the diagonal entries
in the Lagrangian \eqref{lagrangian}. Kinetic~\cite{Holdom:1985ag} and magnetic mixing~\cite{Bruemmer:2009ky}
between the different U(1) sectors is represented by the off-diagonal $\chi$-term and $\theta_{\chi}$-term, respectively.

It should be noted that both the kinetic mixing and the magnetic mixing term naturally arise via quantum corrections
when heavy particles are integrated out~\cite{Holdom:1985ag,Bruemmer:2009ky} even when both U(1)s arise from a single non-abelian gauge group~\cite{Bruemmer:2009ky}.

Let us now see how minicharges can arise in this setup by looking at two simple examples.\\
{\bf 1. Kinetic mixing:} Consider a situation in which all the $\theta$-terms vanish but the kinetic mixing term $\chi\neq 0$.
The kinetic terms in Eq.~\eqref{lagrangian} can be diagonalized by a shift of the hidden gauge field,
\begin{equation}
\label{shift}
X^{\mu}\rightarrow X^{\mu}-\frac{\chi e_{h}}{e} A^{\mu}.
\end{equation}
Apart from a multiplicative renormalization of the gauge coupling,
\begin{equation}
\label{rescaling}
e^{2}\rightarrow e^{2}/(1-\chi^2),
\end{equation}
the ordinary electromagnetic gauge field $A^{\mu}$ remains unaffected by this shift.
Consider now, for example, a hidden fermion $f$ charged under $X^{\mu}$.
Applying the shift~\eqref{shift} to the coupling term, we find:
\begin{equation}
\bar{f}\Bslash\, f\rightarrow \bar{f}\Bslash\, f-\frac{\chi e_{h}}{e}\bar{f}\Aslash\,
 f.
\end{equation}
Since the kinetic term is now diagonal, it is clear that the particle $f$ (which was originally charged
only under U(1)$_{h}$) interacts with the U(1)$_{\rm QED}$ gauge field and appears to have a charge $\epsilon=-\chi e_{h}/e$.

\noindent {\bf 2. Magnetic mixing:} Let us now look at the effect of the $\theta$-terms. For simplicity let us take $\chi=\theta_{\rm QED}=\theta_{h}=0$,
but $\theta_{\chi}\neq 0$.

Let us assume we have a magnetic monopole under the hidden gauge group.
In a static situation with such a hidden monopole as a background the
electric and magnetic fields of U(1)$_{\rm QED}$ and U(1)$_{h}$ read
\begin{equation}
{\bf E}=\nabla A^{0}
,\qquad {\bf B}=\nabla\times{\bf A},\qquad{\bf E}_{h}=\nabla X^{0}
,\qquad {\bf B}_{h}=\nabla\times{\bf X}+\frac{e_{h} g_{h}}{4\pi}\frac{{\bf r}}{r^3}.
\end{equation}
The magnetic mixing part of the Lagrangian then reads
\begin{eqnarray}
L_{\theta_{\chi}}\!\!&=&\!\!\frac{\theta_{\chi}}{8\pi^2}\int d^{3}x\,({\bf E}\cdot{\bf B}_{h}+{\bf E}_{h}\cdot{\bf B})
=-\frac{\theta_{\chi}}{8\pi^2}\int d^{3}x\, A^{0}\nabla\cdot\left(\frac{g_{h}e_{h}}{4\pi} \frac{{\bf r}}{r^3}\right)
\\\nonumber
\!\!&&\!\!\qquad\qquad\qquad\qquad\qquad\qquad=-\frac{\theta_{\chi}e_{h}g_{h}}{8\pi^2}\int d^{3} x\, A^{0}\delta^{3}({\bf r})
=-\frac{\theta_{\chi}}{2\pi}\int d^{3} x\, A^{0}\delta^{3}({\bf r}),
\end{eqnarray}
where we have used the quantization condition $e_{h}g_{h}=4\pi$ in the last step (see next section).
The last part is, however, exactly what we would expect for the interaction of a particle with an (ordinary) \emph{electric} charge\footnote{This is a generalization
of the Witten effect~\cite{Witten:1979ey}.}
$\epsilon=-\theta_{\chi}/(2\pi)$.

\section{Minicharges and charge quantization}
In the previous section we have seen that minicharges can arise quite naturally in extensions of the Standard Model with an extra U(1) gauge factor.
However, one might wonder how this fits with the idea of charge quantization (for more details see \cite{Brummer:2009cs}).

In the following we will see that the extra U(1) gauge boson plays an important role in the compatibility of minicharges and charge
quantization arguments. In the main text we will focus on a variant of the charge quantization
arguments based on angular momentum quantization~\cite{Wilson:1949} in the presence of monopoles.
In the Appendix~\ref{alternative} we look at the original Dirac argument~\cite{Dirac:1931kp}
and an argument based on the absence of black hole remnants (and which doesn't require magnetic monopoles).

Let us consider a configuration of a magnetic monopole with coupling $g$ and an electric charge $q$ with coupling strength $qe$.
The field angular momentum of such a configuration is
\begin{equation}
\mathbf{L}=\int d^{3}x\,\, \mathbf{x}\times(\mathbf{E}\times\mathbf{B})=\frac{qeg}{4\pi} \hat{\mathbf{n}}.
\end{equation}
Here, $\hat{\mathbf{n}}$ is the unit vector pointing from the electric charge to the magnetic charge, and the right hand side can be obtained
by inserting the electric fields $\mathbf{E}=qe\mathbf{r}/(4\pi r^3)$ and the magnetic field $\mathbf{B}=g\mathbf{r}/(4\pi r^3)$ for the electric and
magnetic monopole, respectively.
The quantization of angular momentum in quantum mechanics now requires
\begin{equation}
\label{anguquant}
|\mathbf{L}|=\frac{qeg}{4\pi}=\frac{n}{2},
\end{equation}
where $n$ is an integer (and as usual $\hbar=1$).
Enforcing Eq.~\eqref{anguquant} for our minicharged particle as well as an electron now requires that the ratio between the charges is rational, i.e. $n/m$ with
$n,m\in\mathbb{Z}$.
This forbids irrational minicharges such as say $\sqrt{2}\times10^{-12}$. However, in \eqref{lagrangian} the parameters
$\chi$ and $\theta_{\chi}$ are arbitrary real numbers. How can this be reconciled?
The answer lies in the extra U(1).

So far we have considered only the ordinary electromagnetic field. However, if the minicharge arises as in the previous
section via kinetic mixing\footnote{Here, we will focus on kinetic mixing. For minicharges arising from magnetic mixing one can follow a similar strategy.}
there is an additional U(1) gauge field. This field can now, too, contribute to the angular momentum.
After the shift~\eqref{shift} the kinetic terms
are diagonal and the generalized expression for the angular momentum is straightforward:
\begin{equation}
\label{angutilde}
\mathbf{L}=\int d^{3}x\,\, \mathbf{x}\times(\mathbf{E}\times\mathbf{B}+\mathbf{E}_{h}\times \mathbf{B}_{h}).
\end{equation}

With two U(1) factors we can, of course, also have more general magnetic monopoles. In general a monopole can have charges
$(g,g_{h})$ under the visible and hidden magnetic fields. Its magnetic field will then be,
\begin{equation}
\left(
  \begin{array}{l}
    \mathbf{B} \\
    \mathbf{B}_{h} \\
  \end{array}
\right)
=\frac{\mathbf{r}}{4\pi r^3}
\left(
  \begin{array}{l}
    g \\
    g_{h} \\
  \end{array}
\right).
\end{equation}
We can now study static configurations of this monopole with:
\begin{itemize}
\item[a)]{} an ordinary electrically charged particle (charge $q=1$) with a field
\begin{equation}
\left(
  \begin{array}{c}
 \!\!   \mathbf{E} \\
    \mathbf{E}_{h} \\
  \end{array}
\right)=\frac{\mathbf{r}}{4\pi r^3}\left(
                                \begin{array}{c}
                                  e \\
                                  0 \\
                                \end{array}
                              \right)
\end{equation}
and
\item[b)]{} a hidden sector particle (hidden charge $q_{h}=1$) that has acquired
a charge $\epsilon$ under the ordinary electromagnetic field,
\begin{equation}
\left(
  \begin{array}{c}
 \!\!   \mathbf{E} \\
    \mathbf{E}_{h} \\
  \end{array}
\right)=\frac{\mathbf{r}}{4\pi r^3}\left(
                                \begin{array}{c}
                                  \epsilon e \\
                                  e_{h} \\
                                \end{array}
                              \right).
\end{equation}
\end{itemize}

Inserting into Eq.~\eqref{angutilde} we find
\begin{equation}
|\mathbf{L}_{a)}|=\frac{eg}{4\pi},\quad\quad|\mathbf{L}_{b)}|=\frac{\epsilon eg+e_{h}g_{h}}{4\pi}=\frac{-\chi e_{h}g+e_{h}g_{h}}{4\pi}.
\end{equation}
Angular momentum quantization now requires that both configurations have half-integer angular momentum,
\begin{equation}
\label{constraint}
|\mathbf{L}_{a)}|=\frac{eg}{4\pi}=\frac{n}{2}\quad {\rm and}\quad |\mathbf{L}_{b)}|=\frac{\epsilon eg+e_{h} g_{h}}{4\pi}
=\frac{-\chi e_{h}g+e_{h} g_{h}}{4\pi}=\frac{m}{2},
\end{equation}
where $m$ and $n$ are integers.
It is clear that a naive monopole with $g_{h}=0$ causes a problem because this would require $|\mathbf{L}_{a)}|/|\mathbf{L}_{b)}|=\epsilon=n/m$
and therefore $\epsilon$ to be rational. In the previous section we have, however, seen that $\epsilon$ arises directly from a parameter of
the low energy Lagrangian and therefore it is typically not rational.
This is the apparent contradiction produced by introducing both monopoles and minicharged particles.

However, a closer inspection of Eq.~\eqref{constraint} reveals two types of monopoles which will not
cause any such problems for arbitrary $\chi$,
\begin{equation}
\label{allowed}
\left(
  \begin{array}{c}
    g \\
     g_{h} \\
  \end{array}
\right)
=\frac{2\pi m}{e_{h}}\left(
   \begin{array}{c}
     0 \\
     1 \\
   \end{array}
 \right),\quad{\rm and}\quad
\left(
  \begin{array}{c}
    g \\
    g_{h} \\
  \end{array}
\right)
=\frac{2\pi n}{e}\left(
   \begin{array}{c}
     1 \\
     \chi \\
   \end{array}
 \right).
\end{equation}
For the first monopole $|\mathbf{L}_{a)}|=0$ and for the second $|\mathbf{L}_{b)}|=0$.

In a pure U(1) setup we can simply choose the appropriate monopoles for consistency. If the U(1) arise from non-abelien gauge groups
monopoles are unavoidable. When the symmetry is broken down to U(1)s they automatically arise as 't Hooft-Polyakov monopoles~\cite{'tHooft:1974qc}.
However, as it turns out~\cite{Brummer:2009cs} these 't Hooft-Polyakov monopoles have exactly the appropriate charges for consistency.

So far we have seen that minicharges can arise naturally in extensions of the Standard Model. However, we have also seen that the way they are
generated as well as their consistency with charge quantization suggests that they are accompanied by a (massless) U(1) gauge boson.
This is not only of theoretical interest but will lead to interesting experimental tests.
In the following three sections we will present tests for charge quantization that do not rely on the extra U(1) gauge boson before we make
use of this feature in Sect.~\ref{sec:lsw}. Finally, in Sect.~\ref{sec:polarization} we look at an experiment which is actually better suited to
constrain minicharged particles \emph{without} hidden photons, showing that one can experimentally distinguish between a situation with/without a hidden photon.

\section{AC/DC an experiment to search for minicharged particles}
Let us now turn to experimental searches for minicharged particles that employ (strong) electric and magnetic fields.
The first experiment is directly sensitive to minicharged particles and does not rely on the existence of an extra hidden photon.

\begin{figure}[t]
\begin{center}
\includegraphics[angle=0,width=.45\textwidth]{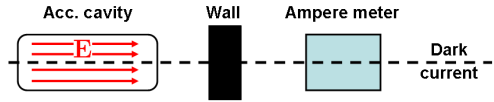}
\end{center}
\vspace{-0.6cm} \caption{\small
Schematic illustration of an
\emph{accelerator cavity dark current} (AC/DC) experiment
 for searching minicharged particles.}
\label{acdc}
\end{figure}

The basic setup for the experiment~\cite{Gies:2006hv} is depicted in Fig. \ref{acdc}.
In a strong electric field a vacuum pair of charged particles gains energy if the particles are separated by a distance along the lines of the electric field.
If the electric field is strong enough (or the distance large enough) the energy gain can overcome the rest mass, i.e. the virtual particles turn into real particles.
This is the famous Schwinger pair production mechanism~\cite{Schwinger:1951nm}.
After their production the electric field accelerates the particles and antiparticles according to their charge in opposite directions.
This leads to an electric current (dashed line in Fig.~\ref{acdc}).
If the current is made up of minicharged particles the individual particles have very small charges and interact
only very weakly with ordinary matter. Therefore, they can pass even through thick walls nearly unhindered. An electron current, however, would be stopped.
After passage through the wall we can then place an ampere meter to detect the minicharged particle current.

\begin{figure}[t]
\begin{center}
\includegraphics[angle=0,width=.6\textwidth]{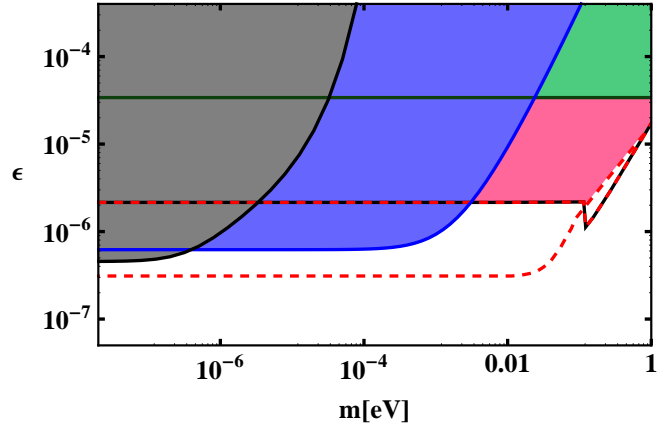}
\end{center}
\vspace{-0cm} \caption{Laboratory bounds on minicharged particles. The black line (on the left, bottom solid) corresponds to the exclusion limit obtained in this
note from the Cavendish type tests of Coulomb's law~\cite{Williams:1971ms,Jaeckel:2009dh}. The blue bound (on the left, middle solid) arises from constraints
on energy losses in high quality accelerator cavities~\cite{Gies:2006hv}.
The dark green curve (on the left, top solid) gives the limit arising from bounds on the invisible
decay on orthopositronium~\cite{Badertscher:2006fm,Ringwald} (a similar bound can be obtained from a reactor experiment~\cite{Gninenko:2006fi}).
The red-black dashed line denotes the limit~\cite{Ahlers:2007rd,Ahlers:2007qf} arising
from light-shining-through-a-wall experiments~\cite{Cameron:1993mr,Ehret:2007cm} and applies only to minicharged particles
arising from kinetic mixing whereas the red dashed curve gives a limit~\cite{Gies:2006ca,Ahlers:2007qf}
from polarization experiments~\cite{Cameron:1993mr,Zavattini:2005tm}$^{4}$ and applies for a pure minicharged particle scenario. The shaded areas are
excluded in both scenarios.}
\label{bounds}
\end{figure}

Typical accelerator cavities achieve field strengths of $\gtrsim 25\,{\rm MeV/m}$ and their size is typically of the order of 10s of cm.
Precision ampere meters can certainly measure currents as small as $\mu {\rm A}$ and even smaller currents of the order of ${\rm pA}$ seem feasible. Using the Schwinger pair production
rate we can then estimate the expected sensitivity for such an experiment to be
\begin{equation}
\epsilon_{\rm sensitivity} \sim 10^{-8}-10^{-6}\quad{\rm for}\,\,m_{\epsilon}\lesssim {\rm meV}.
\end{equation}
Therefore such an experiment has the potential for significant improvement over the currently best
laboratory\footnote{Astrophysical bounds are much stronger~\cite{Davidson:2000hf} but are also somewhat model
dependent~\cite{Masso:2006id}.} bounds shown in Fig.~\ref{bounds}.

\addtocounter{footnote}{1}
\footnotetext[\value{footnote}]{An interesting alternative to polarization experiments is interferometry~\cite{Dobrich:2009kd}.}

Indeed, even without performing a dedicated experiment one can already obtain strong bounds from existing
bounds on the energy loss in accelerator cavities~\cite{Gies:2006hv}. The fact that high $Q$ values have been achieved at large field values strongly
limits the energy loss into unknown sources such as minicharged particle production. The corresponding bound is shown
as the blue region in Fig.~\ref{bounds}.

\section{Tests of Coulomb's law as a probe for minicharged particles}
Minicharged particles would also leave detectable footprints on the distance dependence of the electrostatic potential~\cite{Jaeckel:2009dh}.
In presence of charged particles Coulomb's law is modified due to the effects of the vacuum polarization $\Pi$. This is the so-called
Uehling potential~\cite{Uehling:1935} (for a textbook derivation see, e.g., \cite{Peskin:1995ev}),
\begin{equation}
\label{orig}
V(\mathbf{x})=V_{\rm Coulomb}(x)+\delta V(x)= Q\int\frac{d^{3}q}{(2\pi)^3}\exp(i\mathbf{q}\cdot\mathbf{x})\frac{e^2}{|\mathbf{q}|^2(1-\Pi_{\epsilon}(q))}.
\end{equation}

For large distances, $r\gg 1/m$, $\delta V$ drops off exponentially whereas for small distances, $r\ll1/m$, it's behavior is logarithmic as expected from the
running gauge coupling,
\begin{eqnarray}
\label{Uehlingapprox}
\delta V(r)\!\!&\approx&\!\! \frac{Q\alpha}{r} \left[\frac{\alpha\epsilon^2}{4\sqrt{\pi}}\frac{\exp(-2mr)}{(mr)^{\frac{3}{2}}}\right]
\quad\quad\quad\quad\quad\quad\quad\quad\quad\quad{\rm for}\quad\quad mr\gg 1,
\\\nonumber
\!\!&\approx&\!\! \frac{Q\alpha}{r}\left[-\frac{2\alpha\epsilon^2}{3\pi}\log(2mr)-a\right],
\quad a\approx\frac{2\alpha\epsilon^2}{3\pi}\gamma \quad\quad\,\,{\rm for}\quad\quad mr\ll 1.
\end{eqnarray}

Due to their high mass electrons and heavier charged particles only lead to modifications of Coulomb's law at extremely short distances. In contrast
light minicharged particles could lead to modifications of Coulomb's law at relatively large distances.
We can therefore probe minicharged particles by testing Coulomb's law.

One way to test Coulomb's law is to use a so called Cavendish type experiment. Here, the electric potential between an outer charged conducting sphere
and an uncharged inner sphere is measured\footnote{Of course in reality the experimental setup is often more complicated and
involves multiple spheres~\cite{Williams:1971ms}.}.
A specific feature of Coulomb's law is that the electric potential inside a charged conductor is constant. Therefore, a non-vanishing voltage
between the inner uncharged sphere and the outer sphere would indicate a deviation of Coulomb's law.

By applying high voltages to the outer sphere and looking for minuscle voltages between the inner and the outer sphere
experiments of this type have been performed with enormous precisions of up to about 2 parts in $10^{16}$ for sphere
sizes of 10s of centimeters~\cite{Williams:1971ms}.
This leads to the black bound shown in Fig.~\ref{bounds}.

\section{Tunneling of the 3rd kind -- Searching minicharged particles inside a superconducting box}

\begin{figure}
\begin{center}
\subfigure[]{\scalebox{0.35}[0.35]{
\fcolorbox{white}{white}{
  \begin{picture}(466,194) (42,-121)
    \SetWidth{1.0}
    \SetColor{Black}
    \Photon(48,-14)(160,-14){7.5}{6}
    \GBox(256,-110)(288,82){0.882}
    \Photon(384,-14)(512,-14){7.5}{6}
    \Line[dash,dashsize=10,arrow,arrowpos=0.5,arrowlength=5,arrowwidth=2,arrowinset=0.2](160,-14)(384,-14)
    \Vertex(160,-14){5.657}
    \Vertex(160,-14){7.071}
    \Vertex(384,-14){7.071}
  \end{picture}
}
}
\label{classical}}
\hspace*{0.5cm}
\subfigure[]{
\scalebox{0.35}[0.35]{
\fcolorbox{white}{white}{
  \begin{picture}(466,194) (47,-121)
    \SetWidth{1.0}
    \SetColor{Black}
    \Photon(48,-14)(160,-14){7.5}{6}
    \GBox(256,-110)(288,82){0.882}
    \Arc[clock](208,-14)(48,-90,-270)
    \Line[arrow,arrowpos=0.5,arrowlength=5,arrowwidth=2,arrowinset=0.2](208,34)(336,34)
    \Line[arrow,arrowpos=0.5,arrowlength=5,arrowwidth=2,arrowinset=0.2,flip](208,-62)(336,-62)
    \Arc[clock](336,-14)(48,90,-90)
    \Photon(384,-14)(512,-14){7.5}{6}
  \end{picture}
}
}
\label{t3k}}
\end{center}
\vspace*{-1ex}
\caption{{\bf Left Panel:} Diagram depicting a classical process for a penetration of the
  barrier via conversion into a real particle that interacts only very weakly
  with the barrier. {\bf Right Panel:} Diagram depicting ``tunneling of the 3rd kind''. The photon splits
  into a virtual pair of particle and antiparticle which traverse the wall and
  recombine into a photon.} \label{tunneling}
\end{figure}
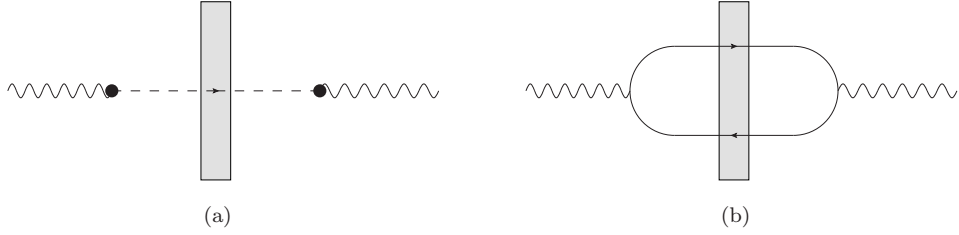

Minicharged particles could also be searched for using a new type of tunneling process, ``tunneling of the 3rd kind''~\cite{Gies:2009wx}.
This process is depicted in Fig.~\ref{t3k}. Here, photons split into a \emph{virtual} particle -- antiparticle pair of charged particles which  traverse
an opaque wall and recombine after the wall into a photon. Due to their tiny interactions with ordinary matter the (virtual) minicharged particles can simply
pass through the wall whereas photons would be stopped.
After recombination the photons can then be detected.
In this way minicharged particles lead to a light-shining-through-a-wall (LSW) signature. In contrast to the
classical LSW process~\cite{regeneration} (Fig.~\ref{classical}) where the photon is converted into a single \emph{real} particle the intermediate particles are \emph{virtual}.
Accordingly, this process depends on the thickness of the wall.

\begin{figure}
\vspace{1cm}
\begin{center}
\includegraphics[width=0.45\linewidth]{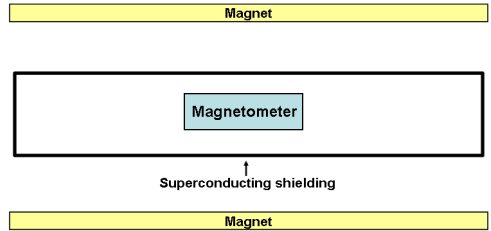}
\end{center}
\vspace{-3ex}
\caption[...]{\small
Sketched setup for the \emph{superconducting box} experiment.
\label{magnet}}
\end{figure}

To observe this process we would like to use a relatively thin wall that is nevertheless completely intransparent to the photons themselves.
For optical frequencies this is rather difficult (among other things due to the high energy of the photons). Therefore, here we will consider
an experiment using static magnetic fields shielded by a superconductor.
The basic idea \cite{Jaeckel:2008sz} of the proposed experiment is very similar to a classic LSW experiments~\cite{regeneration}.
However, instead of light it uses a static magnetic field and the wall is replaced by superconducting shielding (cf. Fig.~\ref{magnet}).
Outside the shielding we have a strong magnetic field. Upon entering the superconductor the ordinary electromagnetic field
is exponentially damped with a length scale given by the London penetration depth $\lambda_{\rm Lon}$.
Yet, due to tunneling of the 3rd kind a small fraction of the magnetic field penetrates the wall and can be detected by a magnetometer.
Since the magnetometer measures directly the field (and not some probability or power output) the signal is proportional
to the transition amplitude and therefore to the charge squared, $\epsilon^2$, instead of being proportional to $\epsilon^4$.

High precision magnetometers can measure
fields of the order of $10^{-13}$~T and even tiny fields of a few~$10^{-18}$~T seem feasible.
The expected sensitivity is of the order of $\epsilon\sim {\rm few}\times 10^{-7}$.

\section{Searching minicharged particles with LSW experiments}\label{sec:lsw}
In the previous section we have already encountered the possibility that minicharged particles can lead to a light-shining-through-a-wall signature.
If there is also a light hidden U(1) gauge boson an additional process is possible and typically dominant: classical LSW via the hidden photon.

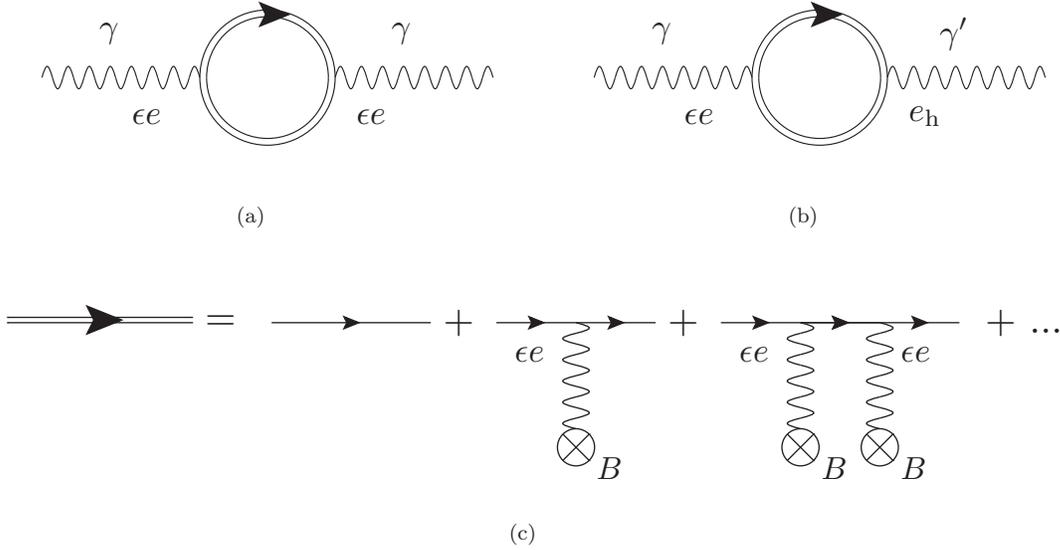
\begin{figure}[t]
\begin{center}
\subfigure[]{
\scalebox{0.85}[0.85]{
\begin{picture}(190,120)(0,40)
\Photon(0,90)(70,90){5}{7.5}
\Text(30,110)[c]{\scalebox{1.5}[1.5]{$\gamma$}}
\Text(160,110)[c]{\scalebox{1.5}[1.5]{$\gamma$}}
\Text(40,73)[l]{\scalebox{1.5}[1.5]{$\epsilon e$}}
\Text(140,73)[l]{\scalebox{1.5}[1.5]{$\epsilon e$}}
\CArc(100,90)(30,0,180)
\CArc(100,90)(30,180,360)
\CArc(100,90)(27,0,180)
\CArc(100,90)(27,180,360)
{\SetWidth{3}
\ArrowLine(100,118)(106,118)}
\SetOffset(0,0)
\Photon(130,90)(200,90){5}{7.5}
\end{picture}
}
\label{convertAA}}
\hspace{1cm}
\subfigure[]{
\scalebox{0.85}[0.85]{
\begin{picture}(190,120)(0,40)
\Photon(0,90)(70,90){5}{7.5}
\Text(30,110)[c]{\scalebox{1.5}[1.5]{$\gamma$}}
\Text(160,110)[c]{\scalebox{1.5}[1.5]{$\gamma'$}}
\Text(40,73)[l]{\scalebox{1.5}[1.5]{$\epsilon e$}}
\Text(140,73)[l]{\scalebox{1.5}[1.5]{$e_\mathrm{h}$}}
\CArc(100,90)(30,0,180)
\CArc(100,90)(30,180,360)
\CArc(100,90)(27,0,180)
\CArc(100,90)(27,180,360)
{\SetWidth{3}
\ArrowLine(100,118)(106,118)}
\SetOffset(0,0)
\Photon(130,90)(200,90){5}{7.5}
\end{picture}
}
\label{convertBB}}
\subfigure[]{\begin{picture}(190,100)(0,20)
\SetScale{1}
\SetOffset(-100,90)
\Line(0,0)(70,00)
\Line(0,2)(70,2)
{\SetWidth{3} \ArrowLine(34,1)(39,1)}
\Text(75,0)[l]{\scalebox{1.5}[1.5]{$=$}}
\SetOffset(00,90)
 \ArrowLine(0,0)(60,0)  \Text(65,0)[l]{\scalebox{1.5}[1.5]{$+$}}
\SetOffset(85,90)
 \ArrowLine(0,0)(30,0) \ArrowLine(30,0)(60,0) \Text(65,0)[l]{\scalebox{1.5}[1.5]{$+$}}
 \Photon(30,0)(30,-40){5}{5} \CArc(30,-47)(7,0,360) \Line(35,-52)(25,-42) \Line(35,-42)(25,-52)
\Text(38,-55)[l]{\scalebox{1.2}[1.2]{$B$}}
\SetOffset(170,90)
 \ArrowLine(0,0)(30,0) \ArrowLine(30,0)(60,0)  \Photon(30,0)(30,-40){5}{5} \CArc(30,-47)(7,0,360)
\Line(35,-52)(25,-42)
 \Line(35,-42)(25,-52)  \Text(38,-55)[l]{\scalebox{1.2}[1.2]{$B$}}
\SetOffset(200,90)
\ArrowLine(0,0)(30,0) \ArrowLine(30,0)(60,0)  \Photon(30,0)(30,-40){5}{5} \CArc(30,-47)(7,0,360)
\Line(35,-52)(25,-42) \Line(35,-42)(25,-52)
\Text(38,-55)[l]{\scalebox{1.2}[1.2]{$B$}}
 \Text(70,0)[l]{\scalebox{1.5}[1.5]{$+\ ...$}}
\SetOffset(0,0)
\Text(92,78)[l]{\scalebox{1.275}[1.275]{{$\epsilon e$}}}
\Text(177,78)[l]{\scalebox{1.275}[1.275]{{$\epsilon e$}}}
\Text(238,78)[l]{\scalebox{1.275}[1.275]{{$\epsilon e$}}}
\end{picture}
\label{convertCC}}
\vspace{.1cm}
\end{center}
\vspace{-0.5cm} \caption[]{\small {
    {The contribution of minicharged particles to the
      polarization tensor \ref{convertAA}. The real part leads to
      birefringence, whereas the imaginary part reflects the absorption
      of photons caused by the production of particle-antiparticle
      pairs.}  The analogous diagram \ref{convertBB} shows how
    minicharged particles mediate transitions between photons and
    hidden-sector photons $\gamma^{\prime}$. Note that the latter diagram is enhanced with respect
    to the first one by a factor {$\sim e_\mathrm{h}/(\epsilon e){=1/\chi}$}.
    The double line represents the complete propagator of the
    minicharged particle in an external magnetic field $B$
    as displayed in \ref{convertCC}. }}
\label{lsw}
\end{figure}

The essential process is depicted in Fig.~\ref{convertBB}. In presence of a non-vanishing magnetic background field the ordinary
electromagnetic photon can be converted into a hidden photon via a minicharged particle loop~\cite{Ahlers:2007rd}. The hidden photon can then traverse the wall
and be reconverted on the opposite side of the wall via the inverse process. In terms of Fig.~\ref{classical} the minicharged particle-loop
is the (resolved) black dot and the hidden photon the dashed particle line.

Currently a sizable number of LSW experiments~\cite{Cameron:1993mr} are under way~\cite{Ehret:2007cm}.
They employ strong lasers as light sources and recycled magnets from accelerator experiments to provide for the magnetic field.
Their power lies in the fact that they can put in $\gtrsim 10^{20}$ photons per second and are often able to detect $\lesssim 1$ photon per second.

The bounds arising from LSW experiments are shown as the red-black dashed line in Fig.~\ref{bounds}. Let us note again, that this bound
on minicharged particles relies on the presence of an extra hidden U(1) gauge boson. But, as we have seen in the first few sections,
this is quite natural.

\section{Searching minicharged particles with laser polarization experiments}\label{sec:polarization}
Finally, let us turn to laser polarization experiments~\cite{Cameron:1993mr,Zavattini:2005tm}.
In these experiments polarized laser light in shone through a strong magnetic field and changes in
the polarization are searched for. These measurements can be used as sensitive probes of minicharged particles~\cite{Gies:2006ca,Ahlers:2007rd,Ahlers:2007qf}.

\begin{figure}
\begin{center}
\subfigure{\begin{picture}(180,100)(0,0)
\includegraphics[width=6cm]{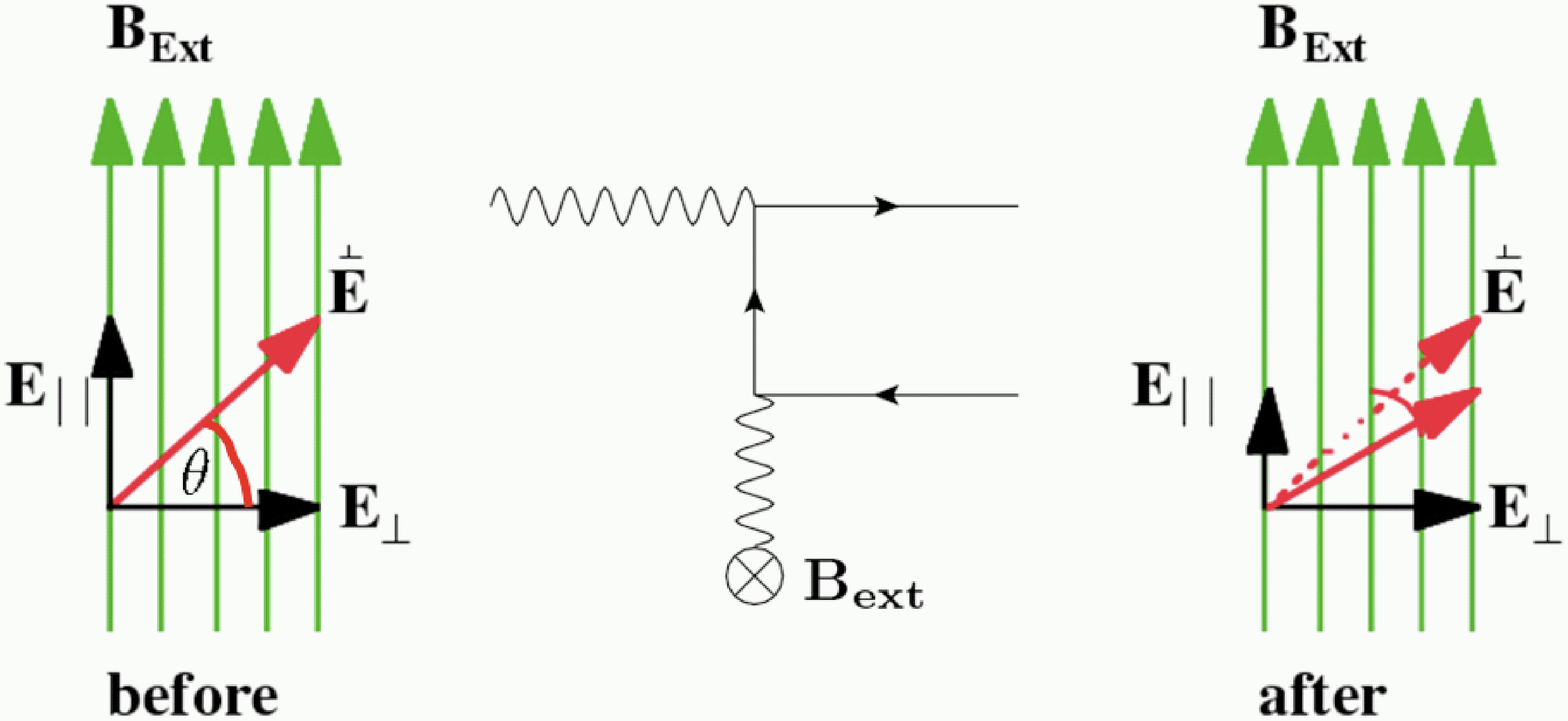}
\end{picture}}
\hspace*{0.5cm}
\subfigure{\begin{picture}(180,100)(0,0)
\includegraphics[width=6.75cm]{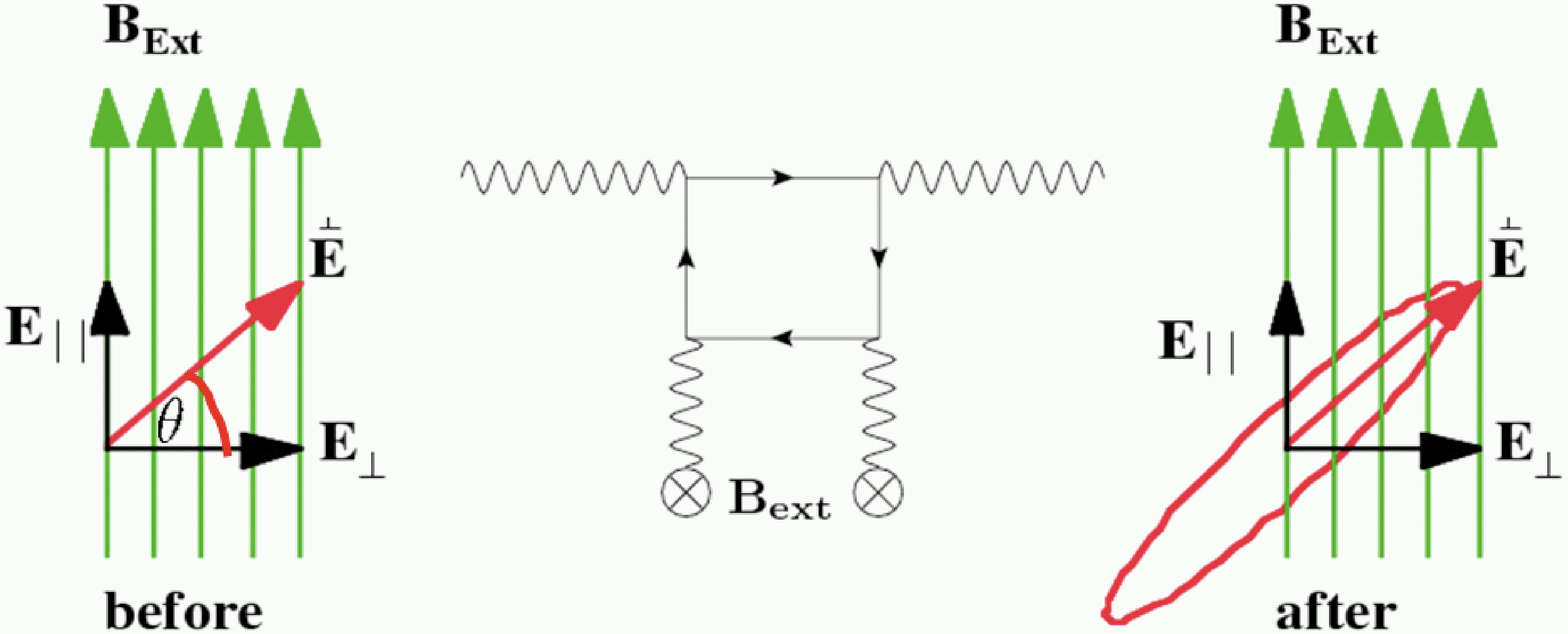}
\end{picture}}
\end{center}
\caption{Rotation (left) and ellipticity (right) caused by the existence of a light
particle with a small electric charge (figure adapted from \cite{Brandi:2000ty}).
More precisely, the Feynman diagram in the left diagram corresponds to the imaginary part of Fig.~\ref{convertAA} and the one in the right
diagram corresponds to the real part of Fig.~\ref{convertAA}.}
\label{polarization}
\end{figure}

We can have two different types of changes in the polarization: a rotation of the linear
polarization and a change into an elliptical polarization.
In particular the rotation is (nearly) free of Standard Model backgrounds and therefore very suitable to probe for new physics~\cite{Ahlers:2008jt}.
As shown in Fig.~\ref{polarization} a selective absorption of one polarization direction\footnote{More precisely, photons of both
polarization directions, parallel and perpendicular to the magnetic field, are absorbed, but the absorption coefficients differ.} via minicharged particle
production results in a
rotation of the linear polarization. A phase shift due to the vacuum polarziation leads to an ellipticity of the polarization.

If there are only minicharged particles Fig.~\ref{polarization} describes the whole story. For this case we obtain the red-dashed bound in Fig.~\ref{bounds}.

In presence of an additional hidden photon the situation is more complicated. Indeed one has to solve the full equations of motion arising from the diagrams
shown in Fig.~\ref{lsw}~\cite{Ahlers:2007rd}. As it turns out in this case the bounds are typically weaker than those obtained with other methods described above.
This can serve as an example that the extra hidden photon can also ``hide'' some of the experimental signatures of minicharged particles.
More positively, this also explicitly shows that we can experimentally distinguish between the two situations.

\section{Conclusions}
Minicharged particles are a natural candidate for physics beyond the Standard Model. Their embedding into extensions of the Standard Model
as well as charge quantization tests suggest that they are accompanied by an extra U(1) gauge boson, the hidden photon.

Light minicharged particles can be searched for in a large variety of laboratory experiments using high precision measurements in strong
electric and magnetic fields. Using different experiments it is possible to distinguish between a situation which only has minicharged
particles and one where the minicharged particle is accompanied by a hidden photon.

It should be mentioned that for minicharged particles astrophysical bounds~\cite{Davidson:2000hf} are often considerably stronger than the current laboratory bounds.
However, astrophysical bounds are also somewhat more model dependent~\cite{Masso:2006id}. Moreover, the recent rapid progress
in the laboratory experiments raises the hope for considerable improvements in the near future.

Finally, the minicharged particles discussed in this note are only one example of a more general class of weakly interacting sub-eV particles (WISPs)
that can be searched for in small scale laboratory experiments at low energies. Most of the above described experiments are also sensitive to other
species of WISPs. For example, the extra hidden U(1) photon could also be massive. This leads to interesting effects such as photon -- hidden photon
oscillations even in absence of extra hidden matter that would become minicharged. For massive hidden photons experiments of the above
described types provide the best bounds -- even stronger than astrophysical bounds -- for a large range of parameters.

With their small scale and rapid development low energy but high precision experiments inside
strong elctromagnetic fields provide an interesting probe of physics beyond
the Standard Model. They are not only complementary in the type of physics they probe but in addition they often recycle parts developed for accelerators
such as high field magnets or high quality cavities.

\section*{Acknowledgements}
The author wishes to thank the organizers of the workshop \emph{Advanced QED Methods for Future Accelerators} for a stimulating meeting.
Moreover he is indebted to S.~Abel, M.~Ahlers, F.~Bruemmer, H.~Gies, V.~V.~Khoze, J.~Redondo and A.~Ringwald for many many interesting
discussions and fruitful collaboration on the subjects discussed in these proceedings.

\begin{appendix}
\section{Alternative charge quantization arguments}\label{alternative}
In this appendix we will briefly consider alternative arguments for charge quantization and investigate how they are compatible with kinetic mixing.
\subsection{A simple variant of the Dirac argument}
In the main text we have used angular momentum quantization to argue in favor of charge quantization. Alternatively we can follow more
closely Dirac's original strategy~\cite{Dirac:1931kp} and look at gauge transformations.

Let us take a look at the gauge field configuration of a Dirac monopole. The gauge field for a Dirac monopole with its string pointing in the
$\mathbf{n}$ direction is,
\begin{equation}
\label{Dirac}
\mathbf{A}^{D}(\mathbf{r})=\frac{g}{4\pi}\frac{\hat{\mathbf{r}}\times\hat{\mathbf{n}}}{r(1-\hat{\mathbf{r}}\cdot\hat{\mathbf{n}})},
\end{equation}
where the hat denotes unit vectors.

A Dirac monopole with its string pointing in the north pole direction $\mathbf{n}_{N}=(0,0,1)$ is physically equivalent to one
with the string pointing in the south pole direction $\mathbf{n}_{S}=(0,0,-1)$.
Therefore, they must be related by a gauge transformation,
\begin{equation}
\label{gt}
\mathbf{A}^{D}_{N}-\mathbf{A}^{D}_{S}=\nabla \Lambda.
\end{equation}
Under this gauge transformation the wave function of a particle with charge $q$ changes according to,
\begin{equation}
\psi\xrightarrow{\Lambda}\exp({\mathbf{i}} q e\Lambda)\psi.
\end{equation}
Since the wave function must be a proper continuous function $\exp({\mathbf{i}} q e \Lambda)$ must be single valued.

Now let us calculate $\Lambda$ for our gauge transformation given by Eq.~\eqref{gt}. Moving the particle from $\mathbf{x}_{0}$ to $\mathbf{x}$ the
gauge phase changes by,
\begin{equation}
\Lambda(x)-\Lambda(\mathbf{x}_{0})=\int^{\mathbf{x}}_{\mathbf{x}_{0}} d\mathbf{s}\nabla \Lambda
=\int^{\mathbf{x}}_{\mathbf{x}_{0}} d\mathbf{s}(\mathbf{A}^{D}_{N}-\mathbf{A}^{D}_{S}).
\end{equation}
Now we can also imagine moving the particle around a closed loop.
The single valuedness of the gauge transformation $\exp({\mathbf{i}} q e\Lambda)$ then requires
\begin{equation}
\label{quant}
2\pi n=q e\oint d\mathbf{s}(\mathbf{A}^{D}_{N}-\mathbf{A}^{D}_{S}), \quad\quad n\in\mathbb{Z}
\end{equation}
for any closed curve.

For our gauge transformation changing the direction of the string from the north to the south pole the right
hand side of Eq.~\eqref{quant} can be easily evaluated for a closed curve around the equator and we obtain the charge quantization condition,
\begin{equation}
2\pi n=q e g.
\end{equation}

In the case of minicharges with kinetic mixing we can now ask ourselves how this quantization condition is fulfilled.
As with the angular momentum argument the trick is to take into account the effect of the hidden magnetic monopole field.
For the ordinary electron which has no hidden electric charge the argument works as above. However, for the minicharged
particle arising from kinetic mixing Eq.~\eqref{quant} picks up an additional contribution from the hidden monopole field,
\begin{eqnarray}
\label{quant2}
2\pi n\!\!&=&\!\!\epsilon e\oint d\mathbf{s}(\mathbf{A}^{D}_{N}-\mathbf{A}^{D}_{S})+ e_{h}\oint d\mathbf{s}(\mathbf{A}^{D,h}_{N}-\mathbf{A}^{D,h}_{S}),
\quad\quad n\in\mathbb{Z}
\\\nonumber \!\!&=&\!\! \epsilon e g+e_{h}g_{h}=-\chi e_{h} g+e_{h}g_{h}.
\end{eqnarray}
This is exactly equivalent to Eq.~\eqref{constraint} and is automatically fulfilled if $g_{h}=\chi g$.

\subsection{Charge quantization from the exclusion of black hole remnants}
A completely different argument for charge quantization can be constructed by requiring that the theory is free of black hole remnants.

Let us assume for simplicity that we have only two types of particles with charge $q_{1}$ and $q_{2}$ with $q_{2}/q_{1}{\not \in}\mathbb{Q}$.
Without restriction we can assume $q_{1}>q_{2}=1$. For simplicity we assume that they have the same mass $m$.
Because the charge ratio is irrational we can make arbitrarily small total charges by combining a number $(n_{1},n_{2})$ or particles (and antiparticles)
\begin{equation}
1\gg Q_{tot}=n_{1}q_{1}+n_{2}q_{2}>0.
\end{equation}
For each $\delta>0$ we can now find a combination such that $|n_{1}|+|n_{2}|=N_{min}(\delta)$ is minimal and $\delta\geq Q_{tot}\neq 0$.
Neglecting the (small) interaction energy the energy of such a configuration is $N_{min}(\delta)m$.

However, on the other hand $N_{min}(\delta)$ depends on $\delta$ and goes to infinity as $\delta\to 0$.
Therefore we can construct a series $\delta_{i}>0$ such that
\begin{equation}
\label{series}
\lim_{i\to\infty}\delta_{i}=0,\quad\quad N_{min}(\delta_{i+1})>N_{min}(\delta_{i})\quad\quad{\rm and}\quad\quad \lim_{i\to\infty}N_{min}(\delta_{i})\to\infty.
\end{equation}
The last statement means that with increasing $i$ the energy of such configurations increases without bound.

Now, let us take such a configuration of charge $Q_{tot,i}\leq\delta_{i}$ and throw it into an uncharged black hole.
Now this black hole can decay by emitting Hawking radiation of
uncharged particles, e.g. photons, until it has reached an extremal Reissner-Nordstrom state\footnote{Quantum gravity effects may already stop
the decay at an earlier stage but this should not essentially change the following argument. We can always look only at sufficiently large
$i$ such that $N_{min}(\delta_{i})m\gg M_{P}$.} with
\begin{equation}
\delta^2_{i}\geq Q^2_{tot, i}=M^{2}_{i}/M^{2}_{p},
\end{equation}
and cannot decay any further without loosing charge. However, for sufficiently large $i$ (cf. Eq.~\eqref{series}), $N_{min}(\delta_{i}) m >M_{i}$ and the black hole
also cannot shed charge by emitting charged particles
because of energy conservation\footnote{Emitting any number of particles smaller than $N_{min}(\delta_{i})$ would only
increase the total charge of the black hole and is therefore forbidden.}.
In other words the created black hole is stable. Indeed we have an infinite number of such remnant states in the theory. This is in violation
of our assumption that the theory is remnant free.

Again we can ask ourselves if minicharges arising from kinetic mixing can circumvent this charge quantization argument. Indeed they can. The reason
is again that the electrically minicharged particles also have a hidden electric charge. A combination of particles with a smaller and smaller visible electric
charge then must have higher and higher hidden charges. This modifies the condition for the charged black hole,
\begin{equation}
M^{2}/M^{2}_{P}\geq Q^{2}+Q^{2}_{h}.
\end{equation}
Due to the high hidden charge the black hole now typically has enough energy/mass to decay directly into charged particles and thereby shed its charge,
allowing for a complete decay.
\end{appendix}
\begin{footnotesize}

\end{footnotesize}


\end{document}